\documentstyle[preprint,aps,epsfig]{revtex}
\begin{document}
\draft
\title{\Large Dielectric Behavior of Nonspherical Cell Suspensions}
\author{Jun Lei$^{1,2}$, Jones T. K. Wan$^1$, K. W. Yu$^1$ and Hong Sun$^{2,3}$}
\address{$^1$Department of Physics, Chinese University of Hong Kong,
 Shatin, NT, Hong Kong}
\address{$^2$Department of Applied Physics, Shanghai Jiao Tong University,
 Shanghai 200 030, China}
\address{$^3$Department of Physics, University of California, Berkeley,
 California 94720-7300, USA}

\maketitle

\begin{abstract}
Recent experiments revealed that the dielectric dispersion spectrum of 
fission yeast cells in a suspension was mainly composed of two 
sub-dispersions. The low-frequency sub-dispersion depended on the cell 
length, whereas the high-frequency one was independent of it. The 
cell shape effect was qualitatively simulated by an ellipsoidal cell 
model. However, the comparison between theory and experiment was 
far from being satisfactory. In an attempt to close up the gap between 
theory and experiment, we considered the more realistic cells of 
spherocylinders, 
i.e., circular cylinders with two hemispherical caps at both ends. 
We have formulated a Green's function formalism for calculating the spectral
representation of cells of finite length.  
The Green's function can be reduced because of the azimuthal symmetry of 
the cell. This simplification enables us to calculate the dispersion
spectrum and hence access the effect of cell structure on the
dielectric behavior of cell suspensions.
\end{abstract}
\vskip 5mm

\section{Introduction}
Dielectric spectroscopy has become a quantitative method of real-time
monitoring of cell growth in suspensions [1--3]. The real-time monitoring has
advantages over conventional techniques and would be important to investigate
the dynamic behavior of cell growth.
There are many factors that may influence the dielectric behavior
of biological materials: structure, orientation of dipoles, surface
conductances, membrane transport processes, etc.  All these factors influence
one another and it is difficult to separate out the effect of a single one.
However, some effects can be dominant at certain ranges of frequencies. For
instance, the dielectric dispersion spectrum of living cell suspensions in
radiofrequencies is mainly determined by the cell shape. The objective of this
work is to develop a theory for such correlation, on which new applications
in biotechnology rely.

More recently, Asami \cite{Asami} reported that the dielectric dispersion 
spectrum of fission yeast cells in a suspension was mainly
composed of two sub-dispersions. The experimental data revealed that the
low-frequency sub-dispersion depended on the cell length, while the
high-frequency one was independent of it. Asami adopted a shell-ellipsoid model
[3], in which an ellipsoid is covered with an insulating shell as the electrical
model of nonspherical biological cells. 
The comparison between model calculation [3] and experimental data \cite{Asami} 
was far from being satisfactory.
Asami suggested that the discrepancy is attributed to the rod-like cell 
shape. However, the depolarization factor needed in his theory is difficult 
to calculate for cells of rod-like shape because of lack of available theories.

In this work, we propose the use of the spectral representation 
\cite{Bergman} for analyzing the cell models. 
The spectral representation is a rigorous mathematical formalism of 
the effective dielectric constant of a two-phase composite material 
\cite{Bergman}. It offers the advantage of the separation of material
parameters (namely the dielectric constant and conductivity) from the cell
structure information, thus simplifying the study. 
From the spectral representation, one can readily derive the dielectric 
dispersion spectrum, with the dispersion strength as well as the 
characteristic frequency being explicitly expressed in terms of the 
structure parameters and the materials parameters of
the cell suspension (see section III below).
The actual shape of the real and imaginary parts of the permittivity 
over the relaxation region can be uniquely determined by the Debye relaxation 
spectrum, parametrized by the characteristic frequencies and the dispersion 
strengths. So, we can study the impact of these parameters on the dispersion 
spectrum directly. 

\section{Formalism}
  \subsection{Spectral representation theory}  
We consider a two-component composite dielectric with complex dielectric 
constant $\epsilon(\bf r)$ equal to 
$\epsilon_2=\varepsilon_2+\sigma_2/j2\pi f$ in the host medium and 
$\epsilon_1=\varepsilon_1+\sigma_1/j2\pi f$ in the embedding medium. 
An interface $\Sigma$ separates the two media.
In a uniform applied electric field ${\bf E}=E_0 \hat{\bf z}$ 
(for convenience, let $E_0=-1$), 
the electrostatic potential satisfies the Laplace's equation:
\begin{equation}
\nabla\cdot [(1-\frac{1}{s}\eta({\bf r}))\nabla\Phi({\bf r})]=0,
\end{equation}
where $s=\epsilon_2/(\epsilon_2-\epsilon_1)$ 
denotes the relevant material parameter and 
$\eta({\bf r})$ is the characteristic function of the 
composite, having value 1 for ${\bf r}$ in the embedding 
medium and 0 otherwise. 
The electric potential $\Phi({\bf r})$ can be solved formally as:
\begin{equation}
\Phi({\bf r})=\Phi_0({\bf r})+\frac{1}{s}\int d{\bf r}^\prime
\eta({\bf r}^\prime)\nabla^\prime G_0({\bf r}-{\bf r}^\prime)
\cdot \nabla^\prime\Phi({\bf r}^\prime)
\end{equation}
where $G_0({\bf r}-{\bf r}^\prime)=1/4\pi|{\bf r}-{\bf r}^\prime|$
is the free space Green's function, 
and $\Phi_0({\bf r})={\bf r}\cdot\hat{\bf z}=z$ is the potential of the 
unperturbed uniform field {\bf E}. It is instructive to convert 
the volume integration into the surface integration via the Green's 
second identity and only deal with the potential on the interface 
$\Sigma$ \cite{Yu2000}. 
We denote an integral-differential operator $\Gamma$:
\begin{equation}
\Gamma\Phi({\bf r})=\oint_{\Sigma}^{\prime}d{\bf S}^\prime
  \cdot \nabla^\prime G_0({\bf r}-{\bf r}^\prime)\Phi({\bf r}^\prime)
  +\frac{1}{2}\Phi({\bf r}), \ \ {\bf r}\in\Sigma,
\end{equation}
to avoid the singularity of $G_0({\bf r}-{\bf r}^\prime)$ 
when the integration variable ${\bf r}^\prime$ approaches the 
point of ${\bf r}$ \cite{Yu2000}.
The integration with a ``prime'' denotes the restricted 
integration which excludes ${\bf r}^\prime={\bf r}$.
Let $\Psi_n({\bf r})$ and $s_n$ be the $n$th eigenfunction 
and eigenvalue of the $\Gamma$ operator respectively.
We can expand $\Phi_0({\bf r})$ and $\Phi({\bf r})$ on the interface 
$\Sigma$ in a series expansion of eigenfunction $\Psi_n({\bf r})$:
\begin{eqnarray}
\Phi_0({\bf r}) & = & \sum_n z_n\Psi_n({\bf r}), \\
\Phi({\bf r}) & = & \sum_n \frac{sz_n}{s-s_n} \Psi_n({\bf r}),
\end{eqnarray}
where $z_n$ are the expansion coefficients.
Then we can write the effective dielectric constant $\bar{\epsilon}$ 
in the Bergman-Milton representation \cite{Bergman}:
\begin{eqnarray}
\bar{\epsilon} & = & -\frac{1}{V}\int dV \epsilon({\bf r}) E_z \\
 & = & \frac{1}{V}\int dV \epsilon_2 [1-\frac{1}{s}
 \eta({\bf r})]\frac{\partial\Phi}{\partial z}\\
 & = & \epsilon_2 \left[1-\frac{1}{V}\sum_n\frac{z_n}
 {s-s_n}\oint_\Sigma d{\bf S} \cdot \hat{\bf z}
 \Phi_n({\bf r})\right]\\
 & = &\epsilon_2 \left[1-p\sum_n\frac{f_n}{s-s_n}\right] ,
\end{eqnarray}
where $p$ is the volume fraction of the suspending cells. Note that
$E_z$ is a dimensionless electric field because $E_0=-1$. 
The eigenvalue $s_n$ and the spectral function $f_n$ can be proved 
to be real and satisfy simple properties that 
$0<s_n<1$, $f_n>0$ and $\sum f_n=1$. 
We shall show that both the spectral function $f_n$ and the eigenvalue 
$s_n$ determine the dielectric behavior of cell suspensions.

  \subsection{Cells with an axis of revolution} 
Now the principal problem is to calculate the eigenvalue $s_n$ 
and the spectral function $f_n$. For many cells interacting 
with one another, it is a formidable task. 
However, in the limit of a dilute cell suspension and weak applied field, 
one can regard the cells in suspension as being noninteracting and 
randomly oriented and the problem is reduced to that of a single cell.
We will consider cells with an axis of revolution, namely,  
the spheroidal and the spherocylinder cells [7] to mimic cells of rod-like
shape.
The prolate spheroid is generated by rotating an ellipse around 
its major axis, while the spherocylinder is obtained by fitting two
hemispherical caps at both ends of a circular cylinder.
For a prolate spheroid, the eigenvalues and eigenfunctions can be 
calculated exactly. The only nonzero $f_n$ equals unity for {\bf E} being 
along the major or minor axis of the prolate spheroid, and the corresponding 
eigenvalues are given by:
\begin{eqnarray}
s_z & = & -\frac{1}{q^2-1}+\frac{q\ln[q+(q^2-1)^{1/2}]}
  {(q^2-1)^{3/2}}, \\
s_x & = & (1-s_z)/2.
\end{eqnarray}
where $q$ is the length to diameter ratio; 
$z$ and $x$ refer to the direction along the major and minor axis 
respectively.
For a spherocylinder, the consideration of the symmetry properties of the 
cell will help us choose the appropriate orthogonal basis for calculating 
the matrix elements of the $\Gamma$ operator. 
Because of the rotation symmetry about the major axis of the spherocylinder, 
the eigenfunction is necessarily of the form 
$(a_n\cos n\theta+b_n\sin n\theta)f(x)$ with $n$ being an integer. 
Due to the inversion symmetry of the cell, 
$f(x)$ must be either odd or even functions. 
It is convenient to expand $f(x)$ in a series of Legendre 
polynomials $P_m(x/l)$, where $2l$ is the length of the cell. 
The applied uniform field {\bf E} can always 
be resolved into two components along the major and minor axes 
of the spherocylinder, so we can calculate the $s_n$ and $f_n$ 
for {\bf E} along the major and minor axes separately.
By symmetry, in order to obtain a nonzero $f_n$, the eigenfunction 
should be the form of $\sum A_m P_{2m+1}(x/l)$ for {\bf E} being
along the major axis, while it reads $\cos\theta\sum B_m P_{2m}(x/l)$ 
for {\bf E} being along the minor axis, with $m=0,1,2,\cdots$.
Using this orthogonal basis, we can calculate a truncated matrix according 
to the precision needed. 
We should remark that the matrix is generally nonsymmetric. 

\section{Dielectric dispersion spectrum}
We show here that from the spectral representation,
one can readily derive the dielectric dispersion spectrum.
Substituting $\epsilon_1=\varepsilon_1 + \sigma_1/j2\pi f$ and
$\epsilon_2=\varepsilon_2 + \sigma_2/j2\pi f$
($\varepsilon$ and $\sigma$ being the real and imaginary parts of
the complex dielectric constant) into Eq.(9),
defining a new parameter $t=\sigma_2/(\sigma_2-\sigma_1)$ and re-defining
$s=\varepsilon_2/(\varepsilon_2-\varepsilon_1)$,
we rewrite the effective dielectric constant $\overline{\epsilon}$
after simple manipulations:
\begin{equation}
\overline{\epsilon}=\epsilon_H + \sum_{n} \frac{\Delta\epsilon_n}
{1+jf/f^{c}_{n}} + \frac{\sigma_L}{j2\pi f}, 
\end{equation}
where $\epsilon_H$ and $\sigma_L$ are the high-frequency dielectric
constant and the low-frequency conductivity respectively, while
$\Delta\epsilon_n$ are the dispersion magnitudes,
$f^{c}_{n}$ are the characteristic frequencies of the $n$th
sub-dispersion.

We have already shown that there are only two poles in the spectral 
representation of the prolate spheroids.
In what follows, we will show that there are two dominant poles in the
spectral representation of the spherocylinder and hence there are two 
sub-dispersions in the dielectric dispersion spectrum. 
The dispersion magnitudes and dispersion frequencies are given by:
\begin{eqnarray}
\Delta\epsilon_1 & = & \frac{1}{3}p\varepsilon_2
  \frac{s_1(s-t)^2}{s(s-s_1)(t-s_1)^2},\\
\Delta\epsilon_2 & = & \frac{2}{3}p\varepsilon_2
  \frac{s_2(t-s)^2}{s(s-s_2)(t-s_2)^2},\\
f^{c}_{1} & = & \frac{\sigma_2s(t-s_1)}{2\pi\varepsilon_2t(s-s_1)},\\
f^{c}_{2} & = & \frac{\sigma_2s(t-s_2)}{2\pi\varepsilon_2t(s-s_2)}.
\end{eqnarray}
Thus, we are able to obtain the dispersion strengths as well as the
characteristic frequencies explicitly in terms of the structure parameters
and the materials parameters of the cell suspension.

The dielectric dispersion spectrum of a dilute suspension of prolate 
spheroids is mainly composed of two sub-dispersions, namely, 
$s_z$ is responsible for the lower frequency one and $s_x$ for the higher one.
For a spherocylinder, we obtain a nonvanishing series of $f_n$ and $s_n$. 
Along the major axis, $f_1$ is dominant for all $q$ and we can omit 
the smaller ones. This dominant $f_1$ is plotted
in Fig.\ref{fig1} against $q$, and the corresponding 
$s_z$ are plotted in Fig.\ref{fig2}, together with the exact result of 
a prolate spheroid. 
As is evident in Fig.\ref{fig1}, we can see that the difference between the 
two models is indeed small. 
Along the minor axis, the solution becomes more complicated. 
The dominant $f_2$ near $q=1$ decreases quickly as $q$ increases; 
another $f_3$ increases and takes over at large $q$. 
These two $f_n$ are also plotted in Fig.\ref{fig1}
and their corresponding eigenvalues are plotted in Fig.\ref{fig3}.
As shown in Fig.\ref{fig3}, the two eigenvalues tend to 
that of a prolate spheroid in the limit of both small and large $q$.

Near $q=2$, the two $f_n$ have comparable values, resulting in two 
sub-dispersions at higher frequency.
These sub-dispersions can interfere with each other, rendering it
difficult to find the characteristic frequencies of the
different sub-dispersions. Physically, the local field is the most
nonuniform in this case.
Nevertheless, we will consider cells of large length and omit this
complication.

With Eqs.(13)--(16), it is easy to calculate the effect of the 
rod-like cell structure on the dispersion spectrum and to compare with 
experiment data.
We will show that the spherocylinder model does give some 
improvement towards the experimental result. 
However, the improvement is too small to close up the gap as Asami expected. 
In fact, we shall see that Asami omitted the material parameters 
which will play an important role in the experimental condition. 
By introducing the conductivity contrast $t=\sigma_2/(\sigma_2-\sigma_1)$, 
we found that a small negative $t$, i.e. $\sigma_1\gg\sigma_2$, should be 
used to close up the discrepancy. 

We estimate $t$ and $s$ by fitting Eqs.(13)--(16) to the experimental
ratio of $\Delta\epsilon_1/\Delta\epsilon_2$ and
$f^{c}_{2}/f^{c}_{1}$, and we get $t=-0.0014$ and $s=5.0$.
It means that $\sigma_1\approx 700 \sigma_2$ and
$\varepsilon_1\approx 0.80 \varepsilon_2$.
The enhanced conductivity of cell cytoplasm is attributed to the membrane
potential. The result is in contrast to the previous (unjustified) claim
that $\sigma_1\approx \sigma_2$.

Table \ref{table1} lists the 
$\Delta\epsilon_1/\Delta\epsilon_2$ ratio and $f^{c}_2/f^{c}_{1}$ 
ratio for both experimental and theoretical results. Using the fitting 
material parameters, the improvement is obvious for both the prolate 
spheroid model and the spherocylinder model, while the difference between 
the two models is quite small.

\section*{Discussion and Conclusion}
In this work, we have applied the spectral representation to the dielectric
dispersion of suspensions of fission yeast cells.
As mentioned by Asami \cite{Asami}, the discrepancies between theory and 
experiment may be attributed to the rod-like cell shape.
For cells of nonconventional shape, however,
there exists no available cell model in the literature and we must     
develop the spectral representation from first principles.

More precisely, we have developed a Green's function formalism [6,8] for
calculating the spectral representation of rods of finite length.
We modelled the rod-like cells as the spherocylinders,
i.e., circular cylinders with two hemispherical caps at both ends.
We solved the spectral representation of the effective dielectric
constant from first principles. 
Similar formalism was adopted for cell suspensions near their 
sub-division point [9--11].

Generally speaking, when the axial ratio $q$ is larger than 4, 
the prolate spheroid model can be employed as a good approximation 
for rod-like cell structures. 
For $q<4$, the dielectric behavior will become more
sensitive to the cell structure of the suspending particles, and 
there are more sub-dispersions than that of the prolate spheroid 
suspensions.

Our model does not include the rotational or vibrational effects,
and our results are expected to be valid only for weak electric fields.
Our model also ignores the multi-shell nature of the cells.
Usually the multi-shell model is used to explain the high-frequency 
steps of spherical cell suspensions. Similar conclusions were found in 
one of our previous paper on multi-shell dielectric spheres in 
electrorheological (ER) fluids \cite{Yu} 
to account for the effects of water coating on the ER effects. 
In Ref.\cite{Yu}, we also showed that the spectral representation 
can still be used for multi-shell model, albeit with a slight
modification.

In Asami's experiment, there exist three subdispersions,
the highest frequency step (above 10MHz) is due to the vacuole
and cell wall as mentioned by Asami, while the two lower
frequency step is evidently dependent on the cell shape. 
And the disperson magnitude of the highest frequency step is much
smaller than that of the two lower frequency ones. So one expects 
that the multi-shell model has only a small effect on the lower
frequency steps. In fact the multi-shell model was used in Asami's
theory, but the discrepancy, as we mentioned in the text, 
is still large.

The large cytoplasmic conductivity is a key result of our investigation.   
We believe that the large cytoplasmic conductivity is reasonable
because the cells have to maintain a higher ion concentration in their
cytoplasm to avoid the shrinkage of cells due to a loss of water across
the cell membrane.
However, to our knowledge, there exists no direct experimental mesurement
on the cytoplasmic conductivity.
In our work, we propose a convenient and practical means of determining
the cytoplasmic conductivity from the dielectric spectroscopy data,
which analysis can be important for biotechnology.

\section*{Acknowledgments}
This work was supported in part by the Direct Grant for Research of the
Research Committee and in part by the Research Grants Council of the
Hong Kong SAR Government.
K.W.Y. acknowledges useful conversation with Professor G. Q. Gu.

\begin{table}[hbp]
\caption{The ratios of the characteristic frequencies 
$\Delta\epsilon_1/\Delta\epsilon_2$ and the ratios of 
the dispersion strengths $f^{c}_{2}/f^{c}_{1}$ listed as a function of
the length to diameter ratio $q$ of the cells. 
The experimental results 
were extracted from Ref.[4] together with the theoretical predictions. 
Both the prolate spheroid model and the spherocylinder model adopt the 
same fitting material parameters determined from the experimental data.
\label{table1}}

\begin{tabular}{ddddddddd}
& & & & & \multicolumn{2}{c}{prolate spheroid}
& \multicolumn{2}{c}{spherocylinder} \\

& \multicolumn{2}{c}{experimental result}
& \multicolumn{2}{c}{Asami theory}
& \multicolumn{2}{c}{model}
& \multicolumn{2}{c}{model} \\
  $q$
& $\Delta\epsilon_1/\Delta\epsilon_2$
& $f^{c}_{2}/f^{c}_{1}$
& $\Delta\epsilon_1/\Delta\epsilon_2$
& $f^{c}_{2}/f^{c}_{1}$
& $\Delta\epsilon_1/\Delta\epsilon_2$
& $f^{c}_{2}/f^{c}_{1}$
& $\Delta\epsilon_1/\Delta\epsilon_2$
& $f^{c}_{2}/f^{c}_{1}$  \\
\tableline

3.46 & 2.22 & 8.95 & 0.900 & 3.00 & 2.26 & 5.34 & 2.77 & 5.89 \\ 
7.17 & 8.65 & 27.4 & 2.07  & 7.73 & 6.10 & 15.3 & 6.67 & 16.4 \\ 
10.2 & 16.4 & 52.6 & 3.39  & 13.0 & 9.94 & 25.9 & 10.5 & 27.2 \\
\end{tabular}
\end{table}

\begin{figure}[hbp]
\caption{The three dominant $f_n$ plotted against the axial ratio $q$:
$f_1$ along the major axis (solid line),
$f_1$ along the minor axis (long dashed line) and
$f_2$ along the minor axis (short dashed line).}
\label{fig1}
\end{figure}

\begin{figure}[hbp]
\caption{The eigenvalue $s_n$ associated with $f_n$ along the major axis
plotted against the axial ratio $q$:
the spherocylinder cell (solid line with filled circles), and
the exact result of the prolate spheroid (solid line).}
\label{fig2}
\end{figure}

\begin{figure}[hbp]
\caption{The eigenvalue $s_n$ associated with $f_n$ along the minor axis
plotted against the axial ratio $q$:
$s_1$ of the spherocylinder cell (solid line with filled squares), 
$s_2$ of the spherocylinder cell (solid line with filled circles), and 
the exact result of the prolate spheroid (solid line).}
\label{fig3}
\end{figure}

\newpage
\centerline{\epsfig{file=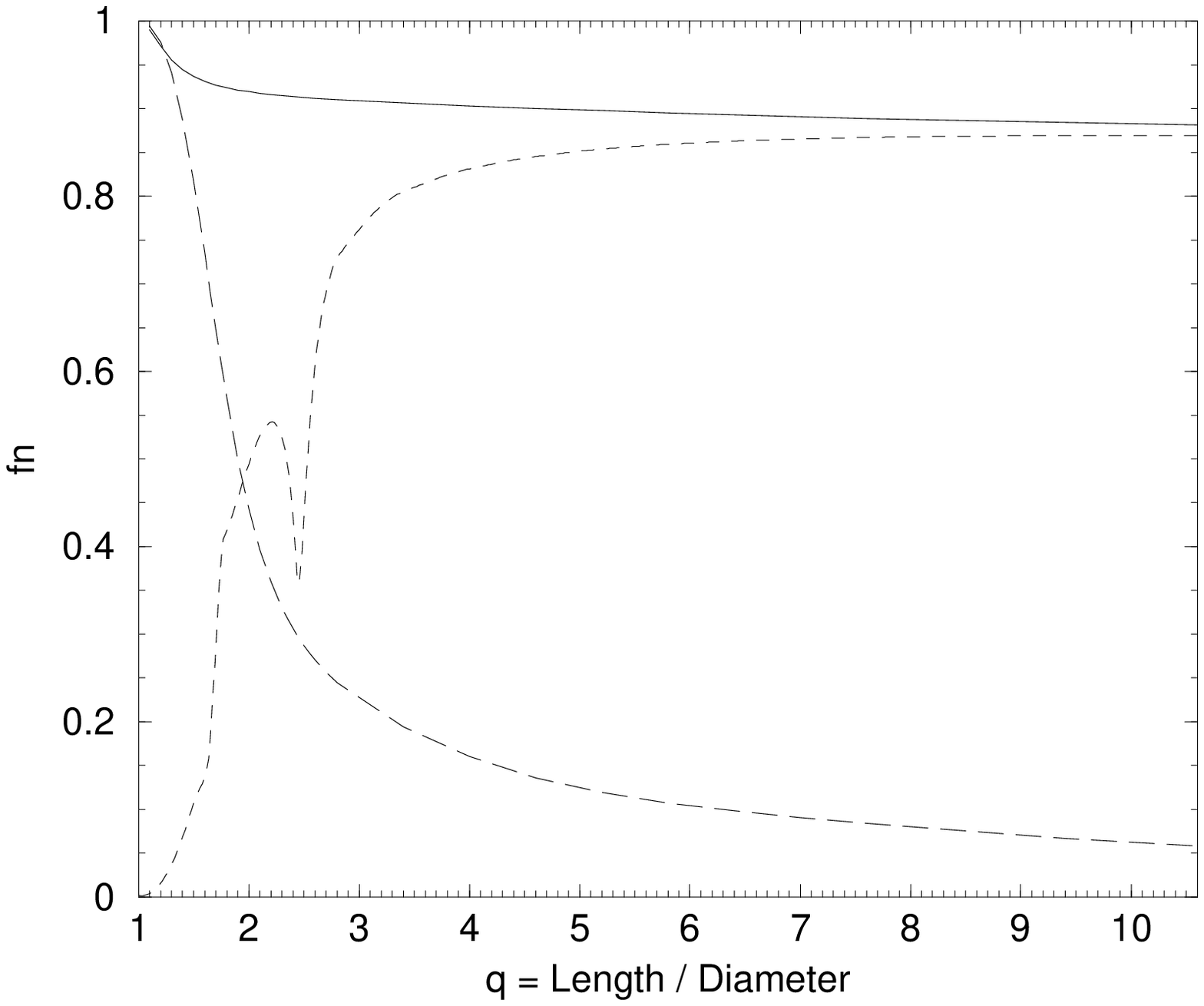,width=\linewidth}}
\centerline{Fig.1/Lei, Wan, Yu, Sun}

\centerline{\epsfig{file=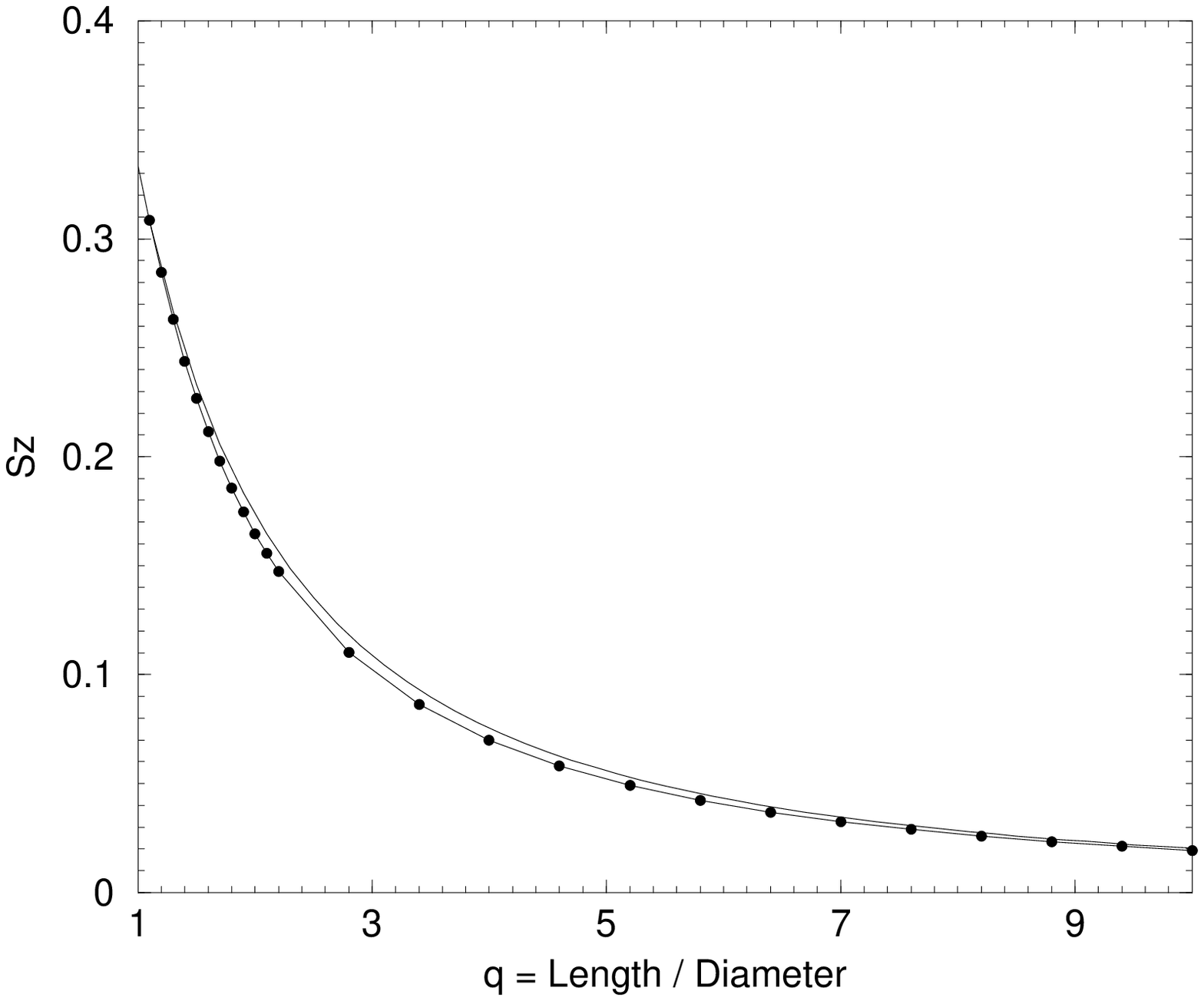,width=\linewidth}}
\centerline{Fig.2/Lei, Wan, Yu, Sun}

\newpage
\centerline{\epsfig{file=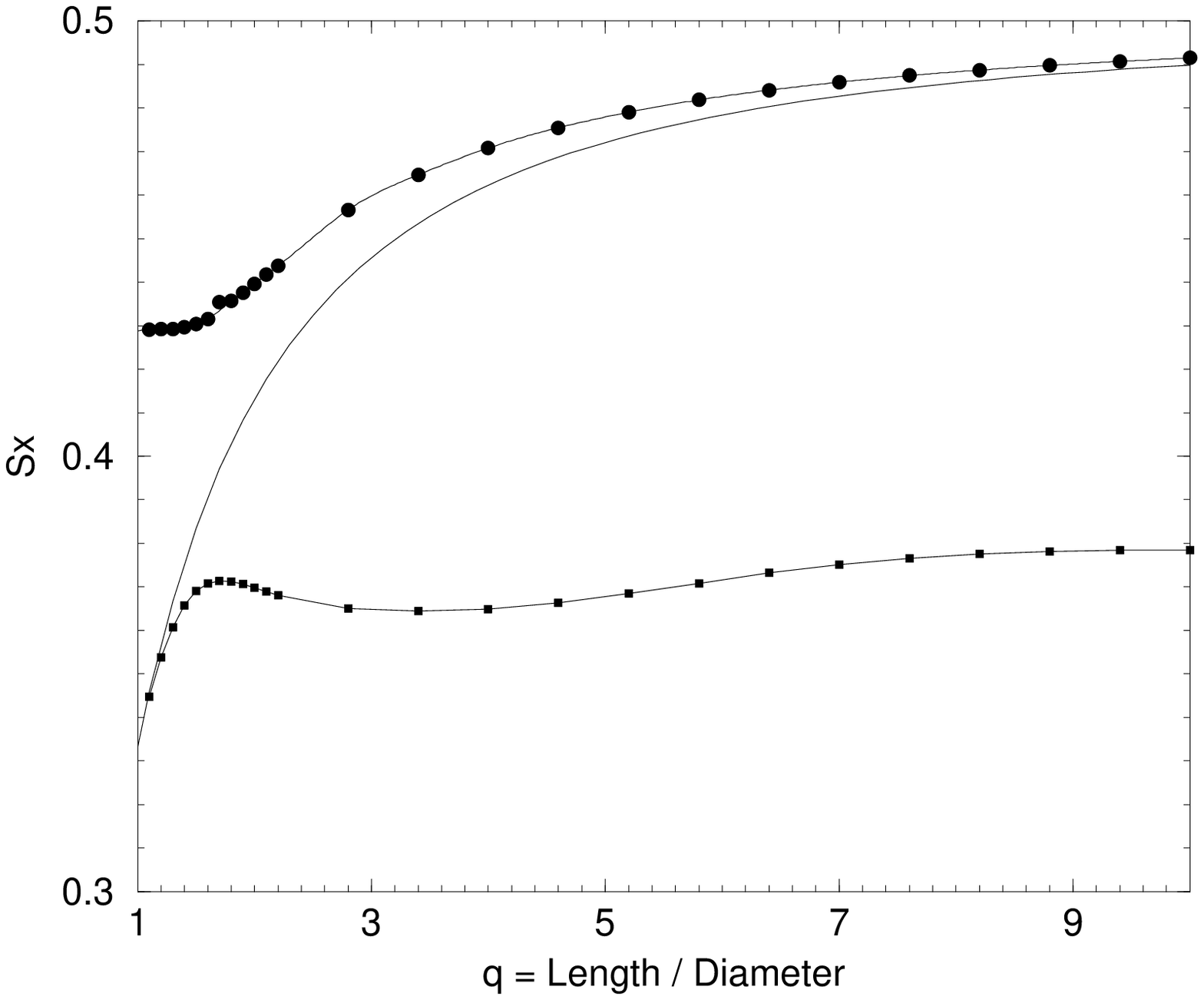,width=\linewidth}}
\centerline{Fig.3/Lei, Wan, Yu, Sun}

\end{document}